\documentclass[letterpaper, english, reprint, aps, superscriptaddress, nofootinbib]{revtex4-1}
\usepackage[T1]{fontenc}
\usepackage[latin9]{inputenc}
\usepackage{amsmath}
\usepackage{amssymb}
\usepackage{physics}
\usepackage{graphicx}
\usepackage{soul}
\usepackage{colortbl}
\usepackage{outline}
\usepackage[colorlinks=true,citecolor=blue,linkcolor=blue]{hyperref}

\usepackage[linesnumbered,ruled]{algorithm2e}

\pdfpageheight\paperheight
\pdfpagewidth\paperwidth

\usepackage{mathdots}
\makeatother

\usepackage{babel}
\begin{document}

% \title{Algorithm Prototyping with Cloud-Based Quantum Computing: \protect\\ A Framework for Quantum Software Developers}
% \title{A Framework for Algorithm Deployment on Cloud-Based Quantum Computers}
\title{A framework for algorithm deployment on cloud-based quantum computers}
% Algorithm Prototyping with Cloud-Based Quantum Computing

\author{Sukin Sim}
\affiliation{%
 Department of Chemistry and Chemical Biology, Harvard University, 12 Oxford Street, \protect\\ Cambridge, MA 02138, USA
}%
\affiliation{%
 Zapata Computing, Inc., 501 Massachusetts Avenue, \protect\\ Cambridge, MA 02139, USA
}%

\author{Yudong Cao}
\affiliation{%
 Zapata Computing, Inc., 501 Massachusetts Avenue, \protect\\ Cambridge, MA 02139, USA
}%

\author{Jonathan Romero}
\affiliation{%
 Department of Chemistry and Chemical Biology, Harvard University, 12 Oxford Street, \protect\\ Cambridge, MA 02138, USA
}%
\affiliation{%
 Zapata Computing, Inc., 501 Massachusetts Avenue, \protect\\ Cambridge, MA 02139, USA
}%

\author{Peter D. Johnson}
\affiliation{%
 Zapata Computing, Inc., 501 Massachusetts Avenue, \protect\\ Cambridge, MA 02139, USA
}%

\author{Al\'{a}n Aspuru-Guzik}%
\email{alan@aspuru.com}
\affiliation{%
 Zapata Computing, Inc., 501 Massachusetts Avenue, \protect\\ Cambridge, MA 02139, USA
}%
\affiliation{%
 Department of Chemistry and Department of Computer Science, University of Toronto, 80 St. George Street, \protect\\ Toronto, ON M5S 3H6, Canada
}%
\affiliation{%
 Canadian Institute for Advanced Research (CIFAR) Senior Fellow, 661 University Avenue, Suite 505, \protect\\ Toronto, ON M5G 1M1, Canada
}%
\affiliation{%
 Vector Institute, 661 University Avenue, Suite 710 \protect\\ Toronto, ON M5G 1M1, Canada
}%

\date{\today}

\begin{abstract}
In recent years, the field of quantum computing has significantly developed in both the improvement of hardware as well as the assembly of various software tools and platforms, including cloud access to quantum devices.
Unfortunately, many of these resources are rapidly changing and thus lack accessibility and stability for robust algorithm prototyping and deployment.
Effectively leveraging the array of hardware and software resources at a higher level, that can adapt to the rapid development of software and hardware, will allow
for further advancement and democratization of quantum technologies to achieve useful computational tasks.
% for democratizing the access to quantum computing software development.
As a way to approach this challenge, we present a flexible, high-level framework called \texttt{algo2qpu} that is well-suited for designing and testing instances of algorithms for near-term quantum computers on the cloud. Algorithms that employ adaptive protocols for optimizations of algorithm parameters can be grouped under the umbrella of ``adaptive hybrid quantum-classical'' (AHQC) algorithms. We demonstrate the utility of \texttt{algo2qpu} for near-term algorithm development by applying the framework to implement proof-of-principle instances of two AHQC algorithms that have applications in quantum chemistry and/or quantum machine learning, namely the quantum autoencoder and the variational quantum classifier, using Rigetti Computing's Forest platform.
\end{abstract}
\maketitle

\section{Introduction\label{sec:intro}}
Significant development in both theory and experiment over the past several decades has positioned quantum computation as a promising technology for various applications, including quantum simulation \cite{Lloyd1996UniversalSimulators,Aspuru-Guzik.Alan2005SimulatedEnergies,Wecker2014Gate-countComputers,Reiher2016ElucidatingComputers}, discrete optimization \cite{Farhi2014,campbell2018applying}, and more recently, machine learning \cite{Romero2017, Wan2017, Cao2017, Farhi2018ClassificationProcessors, Schuld2018Circuit-centricClassifiers, Mitarai2018QuantumLearning, Schuld2018QuantumSpaces, Huggins2018TowardsNetworks,Havlicek2018SupervisedSpaces, Wilson2018QuantumComputers}. 
%\hs{To Yudong: Any others, Yudong?}. 
While certain quantum algorithms \cite{Shor1994AlgorithmsFactoring,Grover1996ASearch,Harrow2009QuantumEquations} guarantee speedups over their classical counterparts (e.g. Shor, Grover, HHL), 
%many of their 
useful realizations will
require fault-tolerant quantum computation.
%assume computations on error-corrected, fault-tolerant machines. 
%Although such devices are currently inaccessible,
Quantum devices supporting fault-tolerant quantum computation have yet to be realized.
In the meantime, significant effort \cite{Peruzzo2014,Farhi2014,OMalley2016} has been invested in leveraging the capabilities of near-term ``noisy intermediate-scale quantum'' (NISQ) devices that can support on the order of $10^2-10^3$ qubits and $10^3$ quantum operations \cite{Preskill2018}. From the algorithmic standpoint, NISQ devices have inspired a class of algorithms that strategically allocate computational tasks between quantum and classical resources, called hybrid quantum-classical (HQC) algorithms \cite{McClean2016TheoryOfHQC}. For clarity, in this work we call a subset of these HQC algorithms that use classical resources to perform optimization of algorithm parameters,
%in parametrized quantum circuits,
adaptive hybrid quantum-classical (AHQC) algorithms\footnote{We use ``adaptive'' as an umbrella term to describe a group of HQC algorithms that use adaptive protocols or optimizations to update algorithm parameters. Consequently, this encompasses HQC algorithms that utilize the variational principle, such as VQE.}. We show a few examples of AHQC algorithms in Table \ref{table:ahqc_algos} that have applications in several areas including chemistry, machine learning, and factoring.
% \pj{(I think "use of variational calculus" is a bit too strong. Maybe "use of the variational principle" instead? But, more broadly, what makes algo2qpu specific to variational algorithms? I could imagine the framework being used, for example, to test performance of different compilers in an FTQC implementation of, say, Shor's algorithm. Plus, I could imagine there being NISQ-ready algorithms which don't invoke the variational principle.)}.
Several of these AHQC algorithms, notably the variational quantum eigensolver (VQE) \cite{Peruzzo2014} and the quantum approximate optimization algorithm (QAOA) \cite{Farhi2014}, have been widely studied, with the VQE algorithm demonstrated using various quantum computing architectures \cite{OMalley2016,Kandala2017Hardware-efficientMagnets,Hempel2018QuantumSimulator}. To continue and accelerate the advancement of quantum computing technologies in both theory and experiment, there is a growing need for software workflows and frameworks that enable organized, rapid testing of algorithms \cite{Zeng2017, McCaskey2018, McCaskey2018softwarex}.

\begin{table*}[]
\bgroup
\def\arraystretch{1.4}
\begin{tabular}{|c|c|c|}
\hline
\textbf{AHQC Algorithm} & \textbf{Goal(s)} & \textbf{Optimization Problem} \\ \hline
\begin{tabular}[c]{@{}c@{}}Variational Quantum \\ Eigensolver (VQE) \cite{Peruzzo2014, McClean2016TheoryOfHQC} \end{tabular} & \begin{tabular}[c]{@{}c@{}}Estimate molecular properties \\ (e.g. energies)\end{tabular} & Minimize expected energy \\ \hline
\begin{tabular}[c]{@{}c@{}}Quantum Approximate \\ Optimization Algorithm (QAOA) \cite{Farhi2014} \end{tabular} & \begin{tabular}[c]{@{}c@{}}Estimate maximum cut of \\ a graph\end{tabular} & Maximize expected cut size \\ \hline
Quantum Autoencoder (QAE) \cite{Romero2017} & \begin{tabular}[c]{@{}c@{}}Design a circuit for compressing \\ a quantum data set\end{tabular} & Maximize average fidelity \\ \hline
\begin{tabular}[c]{@{}c@{}}Quantum Variational Error \\ Corrector (QVECTOR) \cite{Johnson2017} \end{tabular} & \begin{tabular}[c]{@{}c@{}}Design device-tailored quantum \\ error correction scheme \end{tabular} & Maximize average fidelity \\ \hline
\begin{tabular}[c]{@{}c@{}}Variational Quantum \\ Classification \cite{Schuld2018Circuit-centricClassifiers, Schuld2018QuantumSpaces, Havlicek2018SupervisedSpaces} \end{tabular} & \begin{tabular}[c]{@{}c@{}}Find a circuit that  classifies\\ classical data points\end{tabular} & Maximize log likelihood \\ \hline
Variational Quantum Factoring (VQF) \cite{Anschuetz2018} & \begin{tabular}[c]{@{}c@{}}For a given biprime find \\ its prime factors\end{tabular} & \begin{tabular}[c]{@{}c@{}} Minimize quartic boolean \\ polynomial \end{tabular} \\ \hline
\end{tabular}
\egroup
\caption{Examples of adaptive hybrid quantum-classical (AHQC) algorithms. While the objectives of these algorithms may widely vary, each algorithm leverages adaptive protocols to optimize algorithm parameters that (approximately) solve the problem(s)-of-interest.}
\label{table:ahqc_algos}
\end{table*}

Fortunately, over the last few years, various academic and industrial research groups have developed an ecosystem of software tools for simulating and executing quantum circuits, as reviewed in \cite{RyanLaRose2018}. In addition to advanced simulators, some of these platforms, including Rigetti Computing's Forest \cite{RigettiForest} and IBM's Quantum Experience \cite{IBMQX}, provide cloud access to their respective quantum devices. Alongside the rich suite of data tools already available in the cloud, cloud-based quantum computing is expected to become a crucial resource for both research and commercial applications \cite{Mohseni2017,Castelvecchi2017}.

Though numerous studies have already presented experimental demonstrations of quantum algorithms using cloud-based quantum computing \cite{Devitt2016, Dumitrescu2018}\footnote{Please refer to IBM Quantum Experience's Paper page for a comprehensive list of papers that implement experiments using IBM's cloud service.}, there remains a gap between the development of an abstract algorithm and the experimental demonstration of an instance of that algorithm \cite{Chong2017}.
Filling in this gap is particularly important for quantum algorithms of a heuristic nature \cite{Aspuru-Guzik.Alan2005SimulatedEnergies,Farhi2014,Peruzzo2014} as these algorithms rely on rapid prototyping and testing in order to fine-tune them and gauge their feasibility.
% \pj{(I think this is a great set-up. But, I think this "gap" is broader than next sentence would lead you to believe. I think this is a good place to highlight some other issues motivating the value of such a framework. How about including somewhat of a laundry list of things that help fill in this gap?)}
% \pj{Filling in this gap is particularly important for quantum algorithms of a heuristic nature \cite{Aspuru-Guzik.Alan2005SimulatedEnergies,Farhi2014,Peruzzo2014}. 
% Such algorithms rely on rapid prototyping and testing in order to fine-tune them and gauge their feasibility.
% -software for prototyping
% -identify and handle device idiosyncracies which are only revealed through actual implementations}
Furthermore, for general quantum algorithms, significant effort is required to translate the instance of the algorithm, likely realized as abstract, noiseless quantum circuit(s) assuming all-to-all connectivity, to the corresponding lower-level quantum circuits that consider the connectivity and native gate set corresponding to actual devices.
In the age of NISQ devices, this gap is compounded by the noise in the devices, prompting the need for a way to design ``hardware-efficient'' \cite{Kandala2017Hardware-efficientMagnets} algorithmic instances, or circuit(s) that can execute with high fidelity on a given device.
% \pj{Furthermore, while ... , physical implementations of such algorithms have been shown to perform quite differently than simulations \cite{OMalley2016,Kandala2017Hardware-efficientMagnets,Hempel2018QuantumSimulator}.}
Lastly, with growing efforts to build high-quality, reusable packages and platforms, development of reliable abstractions or frameworks to leverage these resources are necessary to achieve and scale up quantum computations for practical applications.

In this work, we introduce a high-level framework for prototyping and deploying AHQC algorithms, called \texttt{algo2qpu}, which provides a systematic workflow from abstract circuits to machine-supported gate-level circuits to execute on either an available simulator or quantum device. We developed \texttt{algo2qpu} with the hope of streamlining the testing of AHQC algorithms. Such testing facilitates algorithm development as well as experimental design. We note that the abstraction of \texttt{algo2qpu} in principle also allows for implementation of algorithms beyond AHQC algorithms, e.g. the iterative phase estimation algorithm.
In the following sections, we describe \texttt{algo2qpu} in greater detail, present its realization using the Forest platform, and apply the infrastructure to execute proof-of-principle instances of two AHQC algorithms, the quantum autoencoder \cite{Romero2017} and the quantum variational classifier \cite{Schuld2018Circuit-centricClassifiers,Havlicek2018SupervisedSpaces,Farhi2018ClassificationProcessors}, on Forest's simulator and quantum processor via the cloud.

\begin{figure}
\includegraphics[width=0.34\textheight]{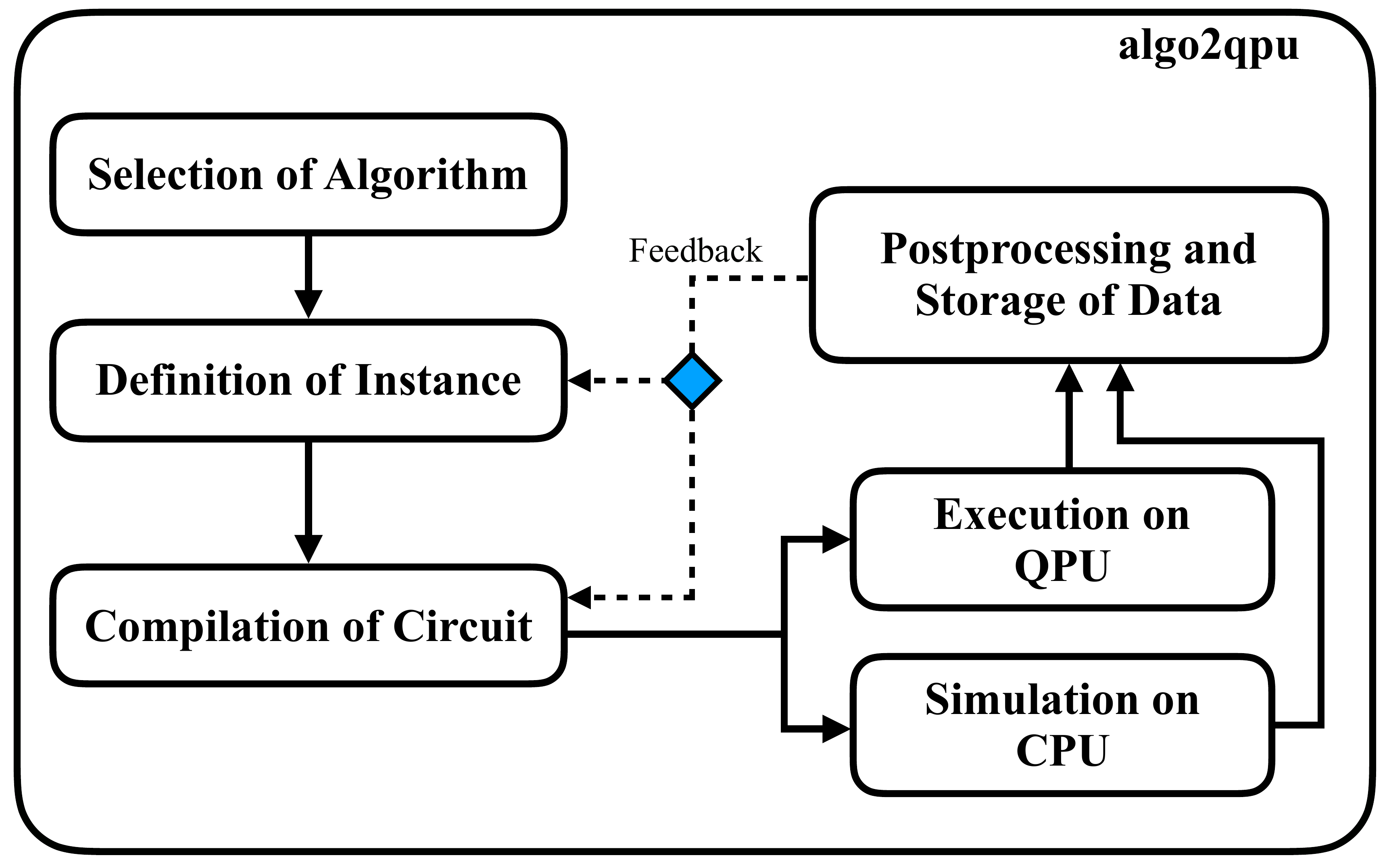}\caption{Overview of \texttt{algo2qpu}: a general high-level framework enabling algorithm-to-instance realizations on cloud-based quantum computers. \label{fig:algo2qpu_general}
}
% \mr{I think it would be helpful to have the box labels all be noun phrases that represent actions (e.g., Selection of Algorithms instead of Algorithm-of-Interest). Also the meaning of the small text within boxes is somewhat inconsistent: an objective function is an object, whereas gate compilation is an action.}
\end{figure}

\section{algo2qpu\label{sec:algo2qpu}}
%\pj{(Should we choose some explicit classical algorithm to follow in order to make the analogy very concrete and easily digestible? Would such an analogy break down at some point in the algo2qpu pipeline?)}
In principle, \texttt{algo2qpu} is a hardware and software agnostic framework well-suited for, but not limited to, implementing instances of AHQC algorithms on a quantum device. This broad framework consists of several major steps: (1) selection of an algorithm-of-interest, (2) definition of a specific instance of the algorithm, (3) compilation of the abstract quantum circuit(s) involved in the algorithmic instance, (4) execution of the compiled circuit(s) to either a simulator or a quantum device, and (5) post-processing and storage of the output data, as shown in Figure \ref{fig:algo2qpu_general}. In the case of AHQC algorithms, a classical feedback routine is integrated to update the algorithm parameters following an adaptive protocol.

% \mr{The list of items is slightly out of alignment with Figure 1: each item corresponds to a box, except for classical feedback, which corresponds to an arrow.}

Below we describe each step in greater detail: \\

\noindent \textit{\textbf{{1. Selection of Algorithm}}} -- 
% \aag{Asked hannah to put table with 3 columns defining HAQC algorithms and also showing examples: column 1 algorithm and citations in brackets, column 2 purpose of algorithm, column 3 objective function to be optimized... useful for non experts. Nobody knows which HAQC algorithms we are really talking about except perhaps US and 3-4 other people around the world!}
At this level, we develop and/or decide on an abstract problem or algorithm to demonstrate. In the case of many AHQC algorithms, this entails defining the objective or cost function(s) that correspond to the particular problem/task. For instance, we may choose to implement VQE, in which the main objective may be to estimate the ground state energy of a fermionic system, that is to minimize the energy of the system. Alternatively, we may be interested in implementing the quantum autoencoder to find a compressed representation of a set of quantum states, or maximizing the fidelity between the input and output data set after applying compression then recovery maps.
% In a programming context (using Pythonic language), this is analogous to defining a class or a function that implements the high-level algorithm. 
% \pj{(Do you think that this step is more analogous to choosing a method for problem solving? Such as choosing a particular convex optimization or linear programming solver?)}
\\

\begin{figure*}
\includegraphics[width=0.92\textwidth]{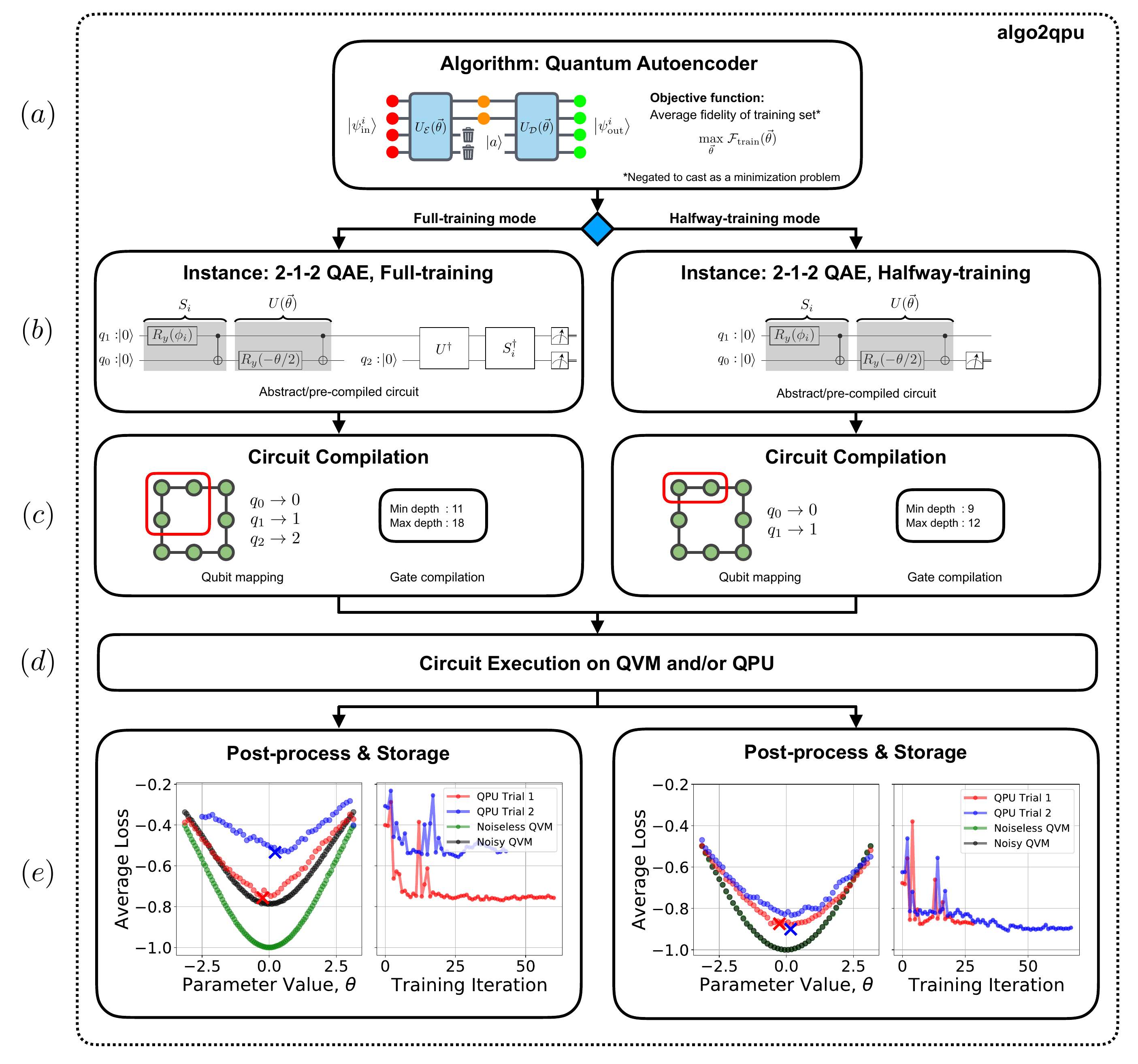}\caption{Quantum autoencoder experiments, applying the full and halfway training techniques, implemented and analyzed using the \texttt{algo2qpu} framework. (a) Schematic representations of quantum autoencoder algorithm for generic quantum circuits. (b) Circuit implementation for the 2-1-2 qubit example presented in the paper. Circuits for state preparation ($S_i$) and compression ($U(\vec{\theta})$) are highlighted. The ``trash'' state is chosen to be $|0\rangle$. (c) Circuit compilation step comprising mapping of the abstract circuits to native gates and qubit index assignment based on connectivity and qubit specifications. We show the minimum and maximum total circuit depth of the compiled circuits along the parameter sweep. (d) Circuit execution step: this involves the simulation (on the QVM) or execution (on the QPU) of the compiled circuits and the classical feedback for carrying out the optimization (not shown). (e) Results of the autoencoder execution: On the left panel for each training mode, parameter sweeps of the average loss for two different executions on the QPU as well as noiseless and noisy simulations performed on the QVM. In addition, we performed two full autoencoder optimization runs on the QPU after each parameter sweep. The results of these optimizations are denoted with crosses (X) in the cost function landscape plots. On the right panel: convergence plots for autoencoder optimization on the QPU, showing the average loss as a function of the number of iterations of the COBYLA optimizer. For clarity, uncertainties in average losses are not displayed. \label{fig:algo2qpu_qae_full_workflow}}
\end{figure*}

\noindent \textit{\textbf{{2. Definition of Instance}}} -- After selecting an algorithm, we define the specific instance of the algorithm in the following step. That is, we consider the quantum circuit(s) used to compute values of the objective function(s) corresponding to the algorithm. In the context of many AHQC algorithms, this generally refers to selecting the parametrized quantum circuits.
% At this stage,
% we \pj{["may or may not"? e.g. IBM's hardware efficient ansatz?]}
% do not assume any knowledge of the quantum device on which to test the algorithmic instance.
At this stage, to consider the circuits at a high level, we label the physical qubits with variables e.g. $q_0$, $q_1$, etc. to indicate that the qubits of the abstract circuits have yet to be assigned to these. (For an example of two instances of a particular algorithm, see Figure \ref{fig:algo2qpu_qae_full_workflow}b).
%we use abstract labeling schemes for the qubits involved, e.g. $q_0$, $q_1$ instead of directly using the physical labels of $0$ and $1$ assigned by the hardware developers, e.g. labels in Figure \ref{fig:agave_connectivity} assigned by Rigetti Computing.
% \jo{I think this example is more confusing than clarifying---why do ``0" and ``1" refer to physical qubits?  Is this a specific reference to the way Rigetti labels qubits?}.
% \pj{(Maybe say that we label the physical qubits with variables $q_0$, $q_1$, etc. to indicate that the qubits of the abstract circuit have yet to be assigned to these.)}
% In a programming context, this could be equivalent to instantiating a class or a function that implements the high-level algorithm by assigning values to all the parameters that are required for describing that instance. 
\\

\noindent \textit{\textbf{3. Compilation of Abstract Circuits}} -- Compilation of quantum circuits according to %limited
device specifications has been an area of research devoted to map a theoretical quantum circuit in the classroom or on paper to a ``lower-level'' quantum circuit whose instructions can be directly processed and executed on a quantum device \cite{Zulehner, Paler2018, Haner2018, Heyfron2017, Nam2017, Venturelli2018, Steiger2018, SABRE_mapping}. We note that the compilation step in the NISQ era can be roughly divided into two major subcomponents: qubit mapping and gate compilation. In addition, this is the step in the workflow that assumes knowledge of the quantum device (e.g. connectivity or single-qubit and two-qubit fidelities). We note that while compilation in today's quantum computing may more closely resemble a simpler, high-level variant,
% or perhaps \textit{transpiling} \jo{Might want to explicitly define this term or give an example since it doesn't seem used very commonly},
significant effort is being devoted to advance compilation such that the process is more analogous to the compilation process for classical computers \cite{JavadiAbhari2014, Chong2017}.

Prior to executing a quantum circuit on an actual device,
%Prior to considering the gates involved in the circuit,
the abstract qubits must be mapped onto the physical qubits of an actual device. This mapping procedure may be non-trivial due to the limited connectivity as well as the quality of the qubits on the processor. That is, even on a single device, there may be a range of qualities or fidelities among the qubits. For example, if the circuit-of-interest requires high quality two-qubit interactions for a subset of qubits, assigning those qubits to physical qubits with high two-qubit fidelities may become a priority in the mapping process.
% \jo{Doesn't it make sense to first consider compiling the gates before choosing physical qubits?  For instance, if I choose some particular assignment, it may be that after gate compilation the physical qubit assignment isn't optimal if the majority of the two-qubit gates are acting between two qubits admitting low fidelities.  On the other hand, optimal gate compilation is probably NP-hard, so choosing physical qubits before hand might be one way to simply that process; is that what you are thinking here?  Geometry.}

% Following the qubit mapping,
In addition, the gates in abstract circuits are generally non-native to a particular device. For example, common gates such as CNOT or Hadamard operations are not native to several existing devices and must be decomposed in terms of the native gates of the chosen platform. We note that several existing compilation routines can also provide valuable circuit information or resource estimates, such as the circuit depth. Depending on the capabilities of the hardware-of-choice, at this stage, one may choose to re-design the experiment to work within the limitations. At the end of the compilation step of \texttt{algo2qpu}, the circuits are ready for execution on a simulator or a quantum device. 
%\pj{(Depending on how we time this with Rigetti's upgrade, we should add something about parametric gate compilation.)}
% \pj{(Also, should we keep up the analogy to classical programming? "This step is analogous to the compiler...")}

While our demonstrations will implement the mapping procedure before the gate compilation, we note that a deeper investigation is necessary to determine the effective ordering of these protocols that correspond to the optimal compilation of the circuit(s).
\\

\noindent \textit{\textbf{4. Circuit Simulation/Execution}} -- Once the quantum circuits are compiled according to the hardware specifications, they can be executed on either a simulator or an actual quantum device (also called the quantum processing unit or QPU) through the cloud. In addition to a wide array of available circuit simulators (e.g. Cirq, QISKit, pyQuil, ProjectQ) \cite{Cirq, Qiskit, RigettiForest, ProjectQ}, some with noise-simulating capabilities, various platforms for cloud-based quantum computing have also been integrated (e.g. Forest and IBM Quantum Experience).
%\jr{I think it would be important to clarify that some of these platforms, most of them actually, are integrated, allowing deployment to either QVM or QPU, and that includes IBM, Google, Rigetti}.
Circuit executions on these services will generally output a collection of measurement outcomes, which can then be post-processed in the following step of \texttt{algo2qpu}. 
% \pj{(Is the analogy job submission, e.g. on AWS?)}
\\

\noindent \textit{\textbf{5. Postprocessing and Storage}} -- The post-processing routine in the framework is responsible for gathering the measurement outcomes and computing the values of the algorithm instance's objective function(s). For NISQ devices, post-processing may significantly benefit from additional error mitigation routines \cite{Endo2017, Temme2017, McClean2017ErrorMitigation, Kandala2018NoiseMitigation}. Both the raw and processed outcome information can also be stored in this step for verifiability. \\

\noindent \textit{\textbf{6. Classical Feedback}} -- For AHQC algorithms, the optimization of algorithm parameters is offloaded to the classical computer. In the context of \texttt{algo2qpu}, once the parameters have been updated by the classical optimizer, they can be updated at the level of the abstract circuit or the compiled circuit that employs parametric variables. Figure \ref{fig:algo2qpu_general} illustrates the possibility of creating a new abstract circuit or directly adjusting the parameters in a compiled circuit in the classical feedback process.
% \mr{Might be good to make Fig. 1 more agnostic about whether feedback involves making a new abstract circuit, a new compiled circuit, or just adjusting variables in an already compiled circuit.}
Then, the re-compiled circuit is executed on the simulator or device until the convergence criteria of the AHQC algorithm are satisfied.
\\

% \hs{Peter, could you possibly write a sentence or two on this?}
% \jo{I noticed that there is no mention of doing tomography or decisions regarding treatment of noise on the machine other than post-processing error mitigation.  I'm not sure, but probably this doesn't fit into the workflow from the perspective of the user since some of these steps would likely be impossible with limited access to a QPU.  However, as noise is a critical factor when thinking about NISQ algorithms, it might be good to incorporate some discussion regarding this into the descriptions somewhere.  For example, maybe the circuit you choose (step 1 or 2) depends on the known noise profile of the device. Or perhaps the circuit compilation (step 3) or post-processing (step 5) may have noise-related components.  Even if it's just a sentence, the perspective may help.}
We note that the abstractness and modularity in the framework in principle can allow for the use of different modules and routines to implement the different steps of \texttt{algo2qpu}, based on the advantages of particular software packages or the availability of specific features. For example, one may use Forest's connections to the simulator and hardware while using ProjectQ's generalized routine for circuit compilation.
In addition, depending on the available features of a platform, \texttt{algo2qpu} can be further optimized to reduce latency by executing certain modules, such as compilation and/or execution, locally or remotely (through the cloud) \cite{RigettiForest}.
In the following section, we present one possible realization of the \texttt{algo2qpu} framework, implemented using features and services supported by the Forest platform\footnote{The version of pyQuil used for the study is \texttt{1.9.0}.}.

% \jr{After reading the description I got the impression that this paper would be really useful for a "classical programmer" that wants to understand how quantum computers can be used, and in general, it would help communicate many of the people in companies, for example, why "quantum specialist" are required to program these machines. Some of these steps might be obvious for many people in the community, but very informative for people outside the "HQC realm". Perhaps it would be worth considering submitting the paper to a journal that can reach a broader audience.}

\begin{figure}
\includegraphics[width=0.43\textwidth]{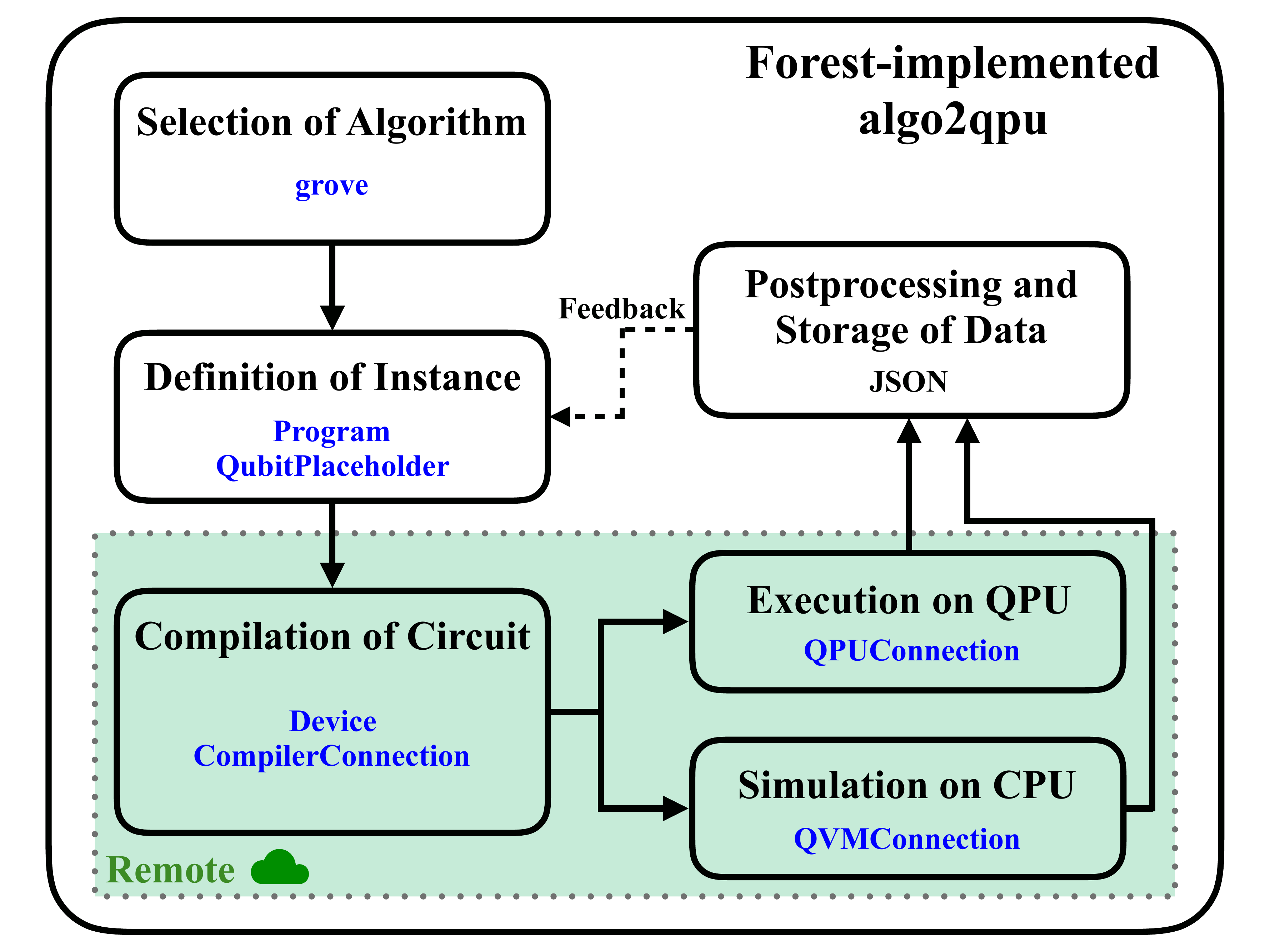}\caption{Forest-tailored \texttt{algo2qpu} framework implemented for this study. Names of Python classes and packages in Forest that were used for implementing aspects of each step of \texttt{algo2qpu} are written in blue. For this particular implementation of the workflow, new abstract circuits with updated parameters are created during the feedback process, and the circuit compilation and execution occur remotely using Forest's cloud service, illustrated using dotted green boxes. \label{fig:forestified_algo2qpu}}
\end{figure}
% \hs{NOTE: Fix names once we have access to Rigetti's updated software! Also, feedback arrow can either point to the ``choose instance'' or ``circuit compilation'' step, given Forest's new parametric compilation feature!} 

\section{Implementation of \texttt{algo2qpu} using Forest\label{sec:algo2qpu_forest}}
\begin{figure}
\includegraphics[height=0.11\textheight]{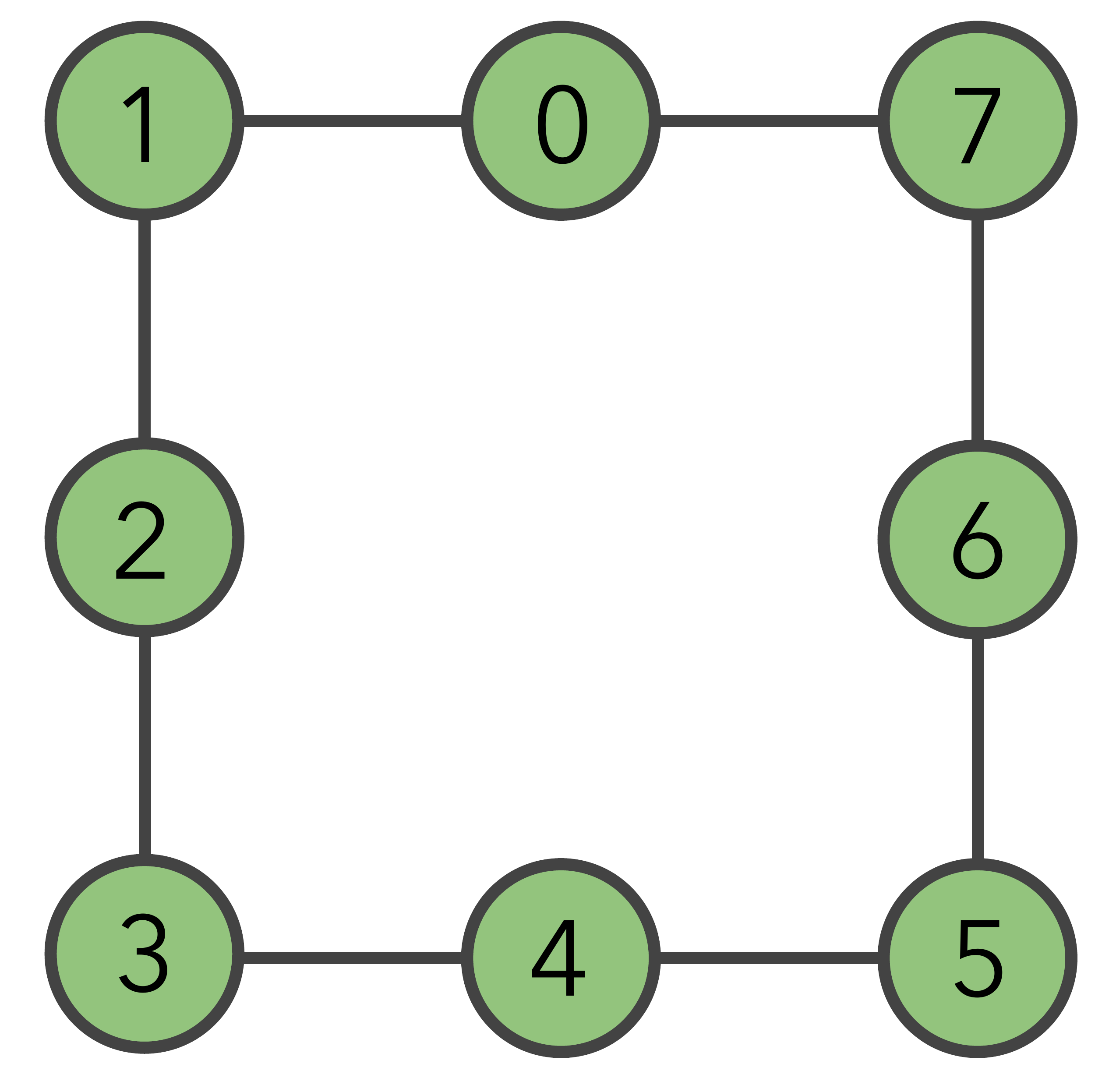}\caption{Device layout for Rigetti Computing's 8Q-Agave quantum processor, with an index associated with each physical qubit. Lines between two qubits indicate the possibility of direct two-qubit interactions. \label{fig:agave_connectivity}}
\end{figure}

% \jo{Anything to say regarding why you chose Forest to demonstrate your example?}
To demonstrate the utility of the \texttt{algo2qpu} framework, we construct each step of the workflow using various modules and/or routines available within the Forest platform, as shown in Figure \ref{fig:forestified_algo2qpu}.
% We chose to implement the framework using Forest because it supports routines and services required by multiple components of \texttt{algo2qpu}.
The choice of algorithm and its instance can be realized by either choosing an algorithmic implementation from grove, Rigetti Computing's library of quantum algorithms, or writing a code using \texttt{pyQuil} that implements an algorithm, such as \cite{QCompress2018}.
The circuit(s) used in the algorithm are represented as instances of \texttt{Program} that employ \texttt{QubitPlaceholder} to label abstract qubits, which can later be replaced with physical qubit indices.
The circuit(s) can be compiled using Forest's in-house compiler that considers the specifications of the particular quantum device.
% qubit mapping directives followed by the use of \texttt{CompilerConnection} for gate compilation that considers the specifications of the particular quantum device, defined by the corresponding instance of the class \texttt{Device}.
At the time of developing \texttt{algo2qpu}, the available quantum device was Rigetti Computing's eight-qubit processor, 8Q-Agave, its layout shown in Figure \ref{fig:agave_connectivity}. Once compiled, the circuit(s) can be executed by connecting to the device and submitting circuit jobs to the cloud, specifically to the quantum virtual machine (QVM) for simulations and the quantum processing unit (QPU) for experiments. The measurement outcomes of both the QVM and QPU are returned as a Python list, which can be post-processed to compute the algorithm's objective(s) and then be stored using the \texttt{json} module.
For a demonstration of the \texttt{algo2qpu} workflow in the context of implementing a specific algorithm, the reader should refer to \cite{QCompress2018}.
%Appendix \ref{app:qae_example}.

In the following sections, we demonstrate the Forest-implemented \texttt{algo2qpu} workflow to design and execute simple instances of two AHQC algorithms: the quantum autoencoder and the variational quantum classifier. For each algorithm, we provide (1) a general description of the algorithm, (2) specific instance(s) of the algorithm, followed by (3) demonstrations and analyses of these instances enabled by \texttt{algo2qpu}.

% \pj{(It would be ideal if we show some lessons learned from the implementation on the QPU. Specifically, how did the implementation of either algorithm using the algo2qpu framework help in improving these algorithms?)}

% \pj{(Kind of picky, but the $q_0$, $q_1$, and $q_2$ are not evenly spaced in (c).)} % fixed
% \jr{Nice figure, a) needs to be modified I think. It might be misleading now because the encoding is acting on the zero state exclusively. Maybe we can make it schematic to match with 4b and the description in sections IV.A,B and C.} 

\section{Algorithm: Quantum Autoencoder\label{sec:qae}}
In this section, we demonstrate simple instances of the quantum autoencoder based on a model proposed by a previous work \cite{Romero2017}. Using \texttt{algo2qpu}, we extend this study by investigating two alternative autoencoder training schemes and providing the first experimental demonstrations of this model on the 8Q-Agave chip.

\subsection{Brief Background\label{subsec:QAE_background}}
The quantum autoencoder (QAE) algorithm has been proposed in recent years \cite{Romero2017, Wan2017} as a paradigm for compressing quantum data, that is expressing a data set comprised of quantum states using a fewer number of qubits. The QAE is constructed using the following: a set of quantum states $\{|\Psi^{i}_{\text{in}}\rangle \}$ of size $n+k$ qubits with $i$ denoting the training set index, a choice of ``trash'' state $|a\rangle$ of size $k$ qubits, $k$ ``refresh'' qubits for the recovery process, and a variational circuit $U(\vec{\theta})$ ($U^{\dagger}(\vec{\theta})$) with circuit parameters $\vec{\theta}$ for encoding (decoding), as illustrated in Figure \ref{fig:algo2qpu_qae_full_workflow}a. In practice, the training set for the QAE has to be prepared on the quantum register using a corresponding set of preparation circuits $\{ S_i \}$ such that $S_i |0\rangle^{\otimes n+k} = |\Psi^{i}_{\text{in}}\rangle$. A successful training of the autoencoder implies finding the parameters that are able to optimally or near optimally factorize all the states in the training set as follows:
\begin{align}
U(\vec{\theta}^{*}) |\Psi^{i}_{\text{in}}\rangle = |\psi_i \rangle \otimes |a\rangle \quad \forall \ i \in [n_{train}],
\end{align}
where $|\psi_i\rangle$ is the compressed representation (on only $n$ qubits) of $|\Psi_i\rangle$ . We can then faithfully recover the input state $|\Psi^{i}_{\text{in}}\rangle $ after applying the decoding operation on the state $|\psi_i\rangle \otimes |a\rangle$, where $|a\rangle$ is prepared on the $k$ ``refresh'' qubits. Consequently, these capabilities may be useful for applications such as dimension reduction of quantum and classical data and feature extraction.

When implementing the autoencoder, we can choose various types of cost functions. In this work, we modify the original autoencoder protocol to accommodate the limitations of the quantum device, avoiding the need for implementing SWAP tests. We first consider a cost function that is computed by executing the full QAE circuit, including, in order, the state preparation, encoding, decoding, and inversed state preparation circuit components. In this case, the success probability of the autoencoder corresponds to the probability of measuring the state $|0\rangle^{\otimes n+k}$ at the end of the circuit. In our implementation, we negate this probability value to cast as a minimization problem and average the frequencies over the training set to compute a single loss value. We will refer to this cost function as ``full-training''. Alternatively, we can consider computing the average probability of measuring the state $|a\rangle$ in the trash register after applying the encoding variational circuit. Similarly, we average over the training set to compute a single loss value. We will refer to this second cost function as ``halfway-training.''

\subsection{Algorithmic Instance: 2-1-2 Autoencoder\label{subsec:QAE_setup}}

To demonstrate our use of the \texttt{algo2qpu} framework for the quantum autoencoder, we consider a simple 2-1-2 instance using both the ``full'' and ``halfway'' autoencoder training schemes, as shown in the left and right panels of Figure \ref{fig:algo2qpu_qae_full_workflow}b. Our data set is composed of two-qubit states, generated by considering a range of $\phi$ values in the state preparation circuit, shown in Figure \ref{fig:algo2qpu_qae_full_workflow}b, in this case forty equally-spaced points from 0 to $\pi$, and the objective is to compress the information such that we can express the input data set using a single qubit. Eight randomly selected points in the data set were selected as the training set. Our simple example is devised such that when the single variational parameter $\theta$ is $0$, this corresponds to the ideal two-to-one compression circuit. We use the notation $q_i$ to refer to qubit indexes in the ``abstract'' QAE circuits in Figure \ref{fig:algo2qpu_qae_full_workflow}b, to anticipate the potentially nontrivial mapping of abstract qubits to physical qubits when implementing the circuit on an actual quantum device with specific connectivity.

% We note that our QAE training circuit construction differs from the circuit originally proposed in \cite{Romero2017}. Instead, we present two new training schemes for the QAE algorithm that are lower in depth, with no additional SWAP tests. We note that the design of this simple instance was based on the capability of the 8Q-Agave chip. In the following subsections, we describe and demonstrate the two training strategies: training of (1) the ``full'' and (2) the "halfway" QAE circuit.

\subsection{Circuit Compilation\label{subsec:QAE_compilation}}
Based on the single- and two- qubit fidelities during the times of the experiments, physical qubits 0, 1, and 2 were selected for the full training scheme, and physical qubits 0 and 1 were selected for the halfway training scheme, with trivial mappings for both cases as shown in Figure \ref{fig:algo2qpu_qae_full_workflow}c. We note that while this selection was done manually for the example in this paper, general workflows will involve automated protocols for qubit mapping. Some examples of these protocols are presented in \cite{Paler2018, Zulehner}. After scanning over values for $\theta$, fifty equally-spaced points from $-\pi$ to $\pi$, we also report the minimum and maximum gate depths for each training scheme in Figure \ref{fig:algo2qpu_qae_full_workflow}c. The gate compilation step was performed using the software tools available in the Forest platform \cite{RigettiForest}.
% Because the halfway training scheme uses only a part of the full QAE circuit, we expect the circuits to execute with higher fidelity. 

\begin{table}[]
\bgroup
\def\arraystretch{1.3}
\begin{tabular}{|c|c|c|}
\hline
\textbf{Setting} & \textbf{\begin{tabular}[c]{@{}c@{}}Training Set\\ Average Loss\end{tabular}} & \textbf{\begin{tabular}[c]{@{}c@{}}Test Set\\ Average Loss\end{tabular}} \\ \hline
Full, QPU (Trial 1) & -0.76 $\pm$ 0.03 & -0.75 $\pm$ 0.02 \\
Full, QPU (Trial 2) & -0.53 $\pm$ 0.05 & -0.44 $\pm$ 0.06 \\
Halfway, QPU (Trial 1) & -0.874 $\pm$ 0.002 & -0.875 $\pm$ 0.002 \\
Halfway, QPU (Trial 2) & -0.90 $\pm$ 0.02 & -0.84 $\pm$ 0.03 \\ \hline
\end{tabular}
\egroup
\caption{Average losses for the training and test sets computed using 8Q-Agave processor applying the two autoencoder training schemes (``full'' for full training and ``halfway'' for halfway training). Standard errors are reported from averaging values over either the training or test set. The minimum loss value is -1.0.}
\label{table:qae_values}
\end{table}

\subsection{Simulation and Experimental Results\label{subsec:QAE_numerics}}
Numerical simulations and experiments of the quantum autoencoder for this study were implemented and executed using an extended version of the \texttt{QCompress} code \cite{QCompress2018}. For both training schemes, each cost function value was evaluated by taking 10000 circuit runs per data point in the training set. For optimizing the variational parameter $\theta$, the COBYLA algorithm was used, with the initial value of $\theta$ set randomly to $\frac{\pi}{1.2} \approx 2.618$ for all experiments.
A parameter sweep for $\theta$ was performed, computing the cost function landscape for fifty equally-spaced points, to assess the impact of experimental conditions on the average loss and on the overall algorithmic performance before each parameter optimization. Two experimental trials were executed for each training scheme, complemented by simulation data \footnote{Tomography experiments are periodically executed by Rigetti Computing to construct a device-imitating noise model for the noisy variant of Forest's circuit simulator.}. 
To evaluate the performance of the autoencoder, we compute the loss values against a test set, which we have pre-selected when randomly splitting the full data set into training and test sets (eight and thirty-two data points respectively).

As shown in Figure \ref{fig:algo2qpu_qae_full_workflow}e, we observe decays in the average loss values for the cost function landscapes of the full and halfway training cases, but the quantum autoencoder was able to reach close to the optimal parameter value of 0 despite the noise in the quantum computer. We also point out that the two executions of the QPU corresponding to full training were performed on different days, and therefore the significant difference might be associated to different calibrations. As expected, due to shorter circuit depths, the cost function landscapes for the halfway training cases better align with results from noisy simulations. In addition, the halfway training scheme produced better average loss values for both training and test sets, as shown in Table \ref{table:qae_values}. This appears to suggest that the halfway training scheme may be a promising alternative and should be further explored with larger instances of the algorithm as a viable training technique for the quantum autoencoder. 
% \jo{We have to be careful here.  The best comparison to make would be the loss on the decoded qubit; comparing a decoded qubit in one case to the trash state on the other is a bit misleading.}
% More generally, the \texttt{algo2qpu} framework allowed for a systematic testing of the QAE instances. That is, we iterated through several instance designs and discarded ones that executed with low fidelity.
\\

\begin{figure*}
\includegraphics[scale=0.23]{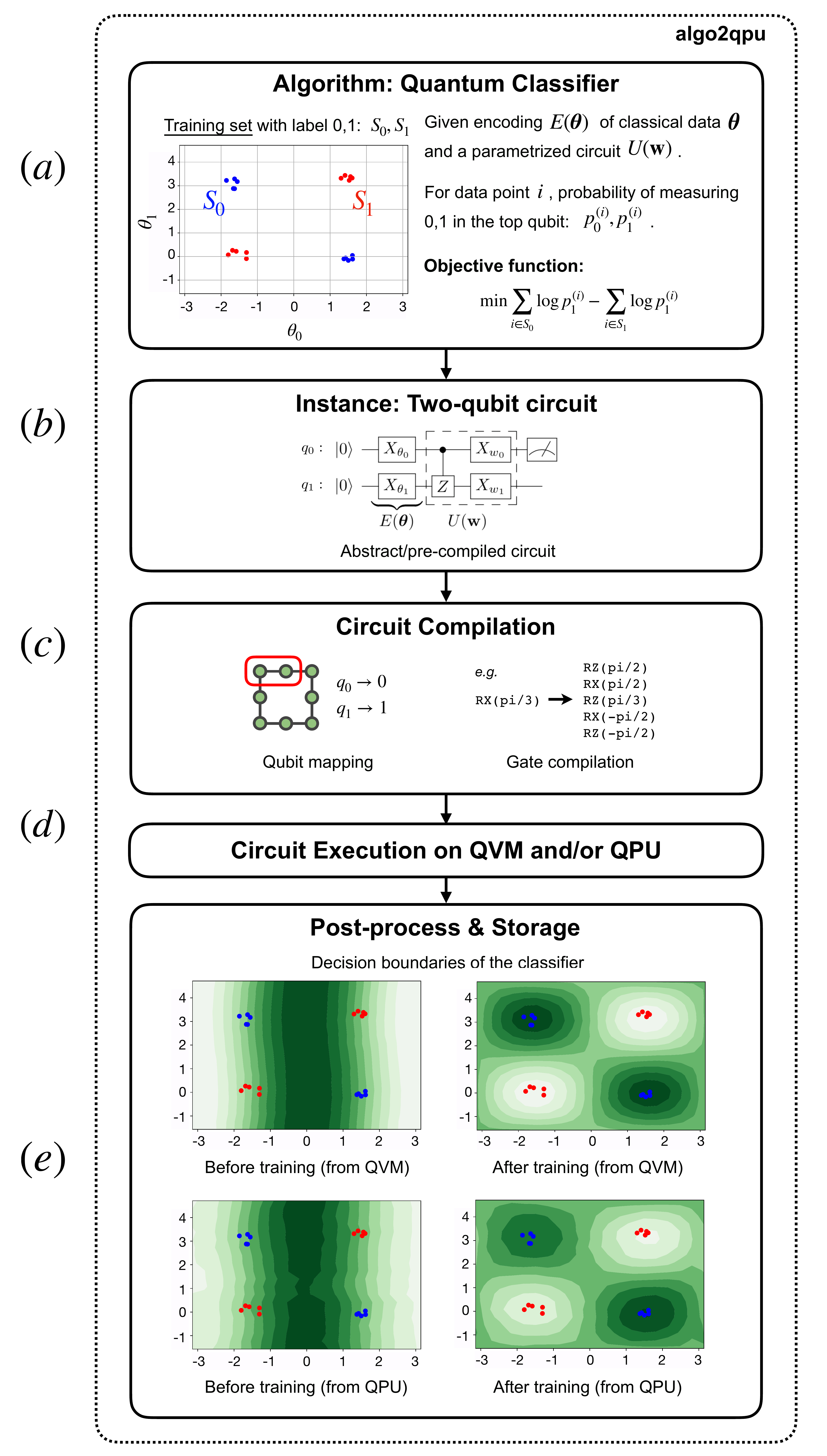}
\caption{Quantum classifier implemented using the {\tt algo2qpu} framework. (a) The basic setting for the classification problem. Here we choose the simplest linearly inseparable data set describing the XOR function. (b) Circuit implementation for a two-qubit classifier. (c) Circuit compilation step comprising mapping of the abstract circuits to native gates and qubit index assignment based on connectivity and qubits specifications. Here the circuit depth is 9 for all parameter assignments. (d) Circuit execution step: this involves the simulation (on the QVM) or execution (on the QPU) of the compiled circuits and the classical feedback for carrying out the optimization (not shown). (e) Contour plot for the probability of measuring $|1\rangle$ in the top qubit (decision boundary of the classifier). Darker color represents higher value. Here we compare decision boundaries of untrained and trained classifiers on both QVM and QPU. Here the untrained classifiers are chosen with the same parameter setting which is specifically intended for them to perform poorly compared with their trained counterparts.}
\label{fig:qnn}
\end{figure*}

\section{Algorithm: Variational Quantum Classifier\label{sec:qclassifier}}
Here we demonstrate a simple example of a variational quantum classifier in full implementation details. The goal here is to provide a minimal example that introduces step by step the workflow of realizing variational quantum classifiers on quantum devices.

\subsection{Brief Background\label{subsec:QClassify_background}}

There has been a rapidly growing set of works in the past few months on using near-term quantum computers for classification problems in machine learning \cite{Farhi2018ClassificationProcessors,Schuld2018Circuit-centricClassifiers,Huggins2018TowardsNetworks,Wilson2018QuantumComputers,Havlicek2018SupervisedSpaces,Schuld2018QuantumSpaces,Mitarai2018QuantumLearning}. Here we consider the problem of binary classification \emph{i.e.\ }learning a function $\mathbb{R}^n\mapsto\{0,1\}$. One of the prevalent methodologies is to variationally tune a quantum circuit and use the measurement outcomes to obtain the output label generated by the quantum model \cite{Farhi2018ClassificationProcessors,Schuld2018Circuit-centricClassifiers,Huggins2018TowardsNetworks,Wilson2018QuantumComputers,Havlicek2018SupervisedSpaces}. For a given training set $S_0\cup S_1$ where $S_0$ consists of all data points labeled 0 and $S_1$ labeled 1, one optimizes the circuit parameter such that for all inputs in $S_0$ the quantum model returns 0 as much as possible and for $S_1$ the model returns 1 as much as possible. There are also alternative proposals for quantum classifiers based on kernel space \cite{Schuld2018QuantumSpaces, Havlicek2018SupervisedSpaces}.
%which are very interesting.
However, here we focus on the variational classifier model.

To implement a variational quantum classifier, one immediate question is how the quantum computer interacts with the classical world. The present literature has more or less reached a consensus on this being a four-part process: (1) encode a classical input vector $\boldsymbol\theta$ into a quantum state $|\Phi(\boldsymbol\theta)\rangle$, (2) apply a variational circuit $U(\bf w)$ of parameters $\bf w$ onto the encoded state, (3) collect measurement statistics of the final state with respect to some operator $M$, (4) classically postprocess the measurement outcomes to obtain the output label $y$ of the quantum model. Different proposals \cite{Farhi2018ClassificationProcessors,Schuld2018Circuit-centricClassifiers,Huggins2018TowardsNetworks,Wilson2018QuantumComputers,Havlicek2018SupervisedSpaces} use different choices for each step, while the overall framework remains largely the same. Here we also adopt this framework in our example.

\subsection{Algorithmic Instance: Learning XOR\label{subsec:QClassify_setup}}

For a set of data points $\boldsymbol\theta_1$, $\cdots$, $\boldsymbol\theta_n$ in $d$ dimensional space, the simplest model for classifying the data points is a linear function %\hs{Describe x} 
$f(\boldsymbol\theta)={\bf w}^T{\boldsymbol\theta}$. Depending on whether the value of the function $f(\boldsymbol\theta)$ is above or below a predefined threshold $b$ we apply a discrete label to the data point $\boldsymbol\theta$.
% Essentially
That is, we are essentially using a hyperplane $f(\boldsymbol\theta)=b$ to separate the data points. The set of $\bf x$ such that $f(\boldsymbol\theta)=b$ 
%\hs{Is there supposed to be some equality here? $f(\boldsymbol\theta) = 0$?} 
therefore forms a \emph{decision boundary} where $\boldsymbol\theta$ is applied one label if $f(\boldsymbol\theta)<0$ \emph{i.e.\ }it lies on one side of the decision boundary and another label if $f(\boldsymbol\theta)>0$ \emph{i.e.\ }it lies on the other side of the decision boundary. 
%\hs{Are these other x's supposed to be theta's?}
% Clearly

If the data points are not linearly separable, then we need to use a different form of $f(\boldsymbol\theta)$ which generates nonlinear boundaries. A classic example of linearly inseparable data is the ``XOR'' dataset\footnote{Historically this is a famed example showing the limitation of perceptrons proposed by Minsky and Papert \cite{Minsky1969}. The fact that XOR functions can be learned by multilayer perceptrons was only widely recognized later.} (Figure \ref{fig:qnn}b). Here we specifically consider an XOR-like training set with $S_0$ being points $\boldsymbol\theta=(\theta_0,\theta_1)$ around $\{(-\pi/2,0),(\pi/2,\pi)\}$, and $S_1$ being points around $\{(-\pi/2,\pi),(\pi/2,0)\}$. 

To encode a classical input $\boldsymbol\theta$, we use the encoding by preparing a quantum state $X_{\theta_0}|0\rangle\otimes X_{\theta_1}|0\rangle$, where $X_\theta = e^{i\frac{\theta}{2}X}$ is a single qubit rotation along the $x$ direction. Similar ideas for encoding classical data have been proposed before \cite{Huggins2018TowardsNetworks}, though in general a more compact encoding is possible \cite{Mitarai2018QuantumLearning,Schuld2018Circuit-centricClassifiers}. We then apply a parametrized unitary $U(\bf w)$ and measure the top qubit. We will use the probability $p_1$ of measuring $|1\rangle$ in the top qubit as the output label (as also considered in \cite{Farhi2018ClassificationProcessors}). The objective function that we would like to minimize is the cross entropy
\begin{equation}
-(\sum_{i\in S_1}\log p_1 - \sum_{i\in S_0}\log p_1).
\end{equation}
In other words, we want to maximize the output $p_1$ (and therefore $\log p_1$) when the data label is 1 and at the same time minimize the output $p_1$ when the data label is 0.

% \begin{figure*}
% \includegraphics[scale=0.23]{algo2qpu-classifier.pdf}
% \caption{Quantum classifier implemented using the {\tt algo2qpu} framework. (a) The basic setting for the classification problem. Here we choose the simplest linearly inseparable data set describing the XOR function. (b) Circuit implementation for a two-qubit classifier. (c) Circuit compilation step comprising mapping of the abstract circuits to native gates and qubit index assignment based on connectivity and qubits specifications. Here the circuit depth is 9 for all parameter assignments. (d) Circuit execution step: this involves the simulation (on the QVM) or execution (on the QPU) of the compiled circuits and the classical feedback for carrying out the optimization (not shown). (e) Contour plot for the probability of measuring $|1\rangle$ in the top qubit (decision boundary of the classifier). Darker color represents higher value. Here we compare decision boundaries of untrained and trained classifiers on both QVM and QPU. Here the untrained classifiers are chosen with the same parameter setting which is specifically intended for them to perform poorly compared with their trained counterparts.}
% \label{fig:qnn}
% \end{figure*}

\subsection{Simulation and Experimental Results\label{subsec:QClassify_numerics}}

We start from an initial guess of the parameter that does not give rise to good classification (Figure \ref{fig:qnn}c) and optimize the circuit parameters using Nelder-Mead. The trained classifier captures closer the XOR function (Figure \ref{fig:qnn}d).

%Acute readers may already have noticed 
Observe that in fact the parameter $w_1$ does not enter in the objective function, rendering the optimization problem as essentially one-dimensional. 
%\hs{Morten: This is a little too colluquial for me. Consider rephrasing to 'Inspection reveals that the parameter w\_1 does not enter in the objective function, reducing the problem to a one-dimensional optimization problem' or something along those lines :) } 
One could in fact absorb the $X_{w_1}$ gate into the measurement of the bottom qubit, and treat $w_1$ as a parameter setting the measurement context only for the bottom qubit. This way the value of $w_1$ does not influence the measurement outcome on the top qubit. We can see this also by explicitly computing the probability $p_1$:
\begin{equation}
\begin{array}{ccl}
p_1 & = & \displaystyle \cos^2\frac{\theta_0}{2}\sin^2\frac{w_0}{2}+\sin^2\frac{\theta_0}{2}\cos^2\frac{w_0}{2} \\[0.1in]
& + & \displaystyle \frac{1}{2}\sin{w_0}\sin{\theta_0}\cos{\theta_1}.
\end{array}
\end{equation}
From the above expression it is clear that the term $\sin(\theta_0)\cos(\theta_1)$ naturally gives rise to the XOR-like landscape in Figure \ref{fig:qnn}d by assigning different signs to its extrema according to the quadrant. The optimal parameter setting is then $w_0=\pi/2$, which gives a $p_1=\frac{1}{2}+\frac{1}{2}\sin(\theta_0)\cos(\theta_1)$ that contains only the desired term.  

% \pj{(PJ: I think that the aim should be to show the value of the algo2qpu framework through these experimental demonstrations. (We have more to offer in this than the group which did all of the ibmqx implementations ;-p). So, I think we should aim to use each implementation to feature algo2qpu: how did the framework help us learn something beyond "the algorithm works in this small instance"? I think there should be some conclusion to draw from this impmlementation.)}

Finally we remark that the observation that the circuit in Figure \ref{fig:qnn}b can be effective in partitioning the XOR data comes from iterative trial and error within the {\tt algo2qpu} framework. Initially, the circuit is trained to classify an XOR-shaped data set similar to Figure \ref{fig:qnn}b but shifted, with $S_0$ being points around $\{(-\pi/2,-\pi/2),(\pi/2,\pi/2)\}$ and $S_1$ around $\{(-\pi/2,\pi/2),(\pi/2,-\pi/2)\}$. From plotting the decision boundary of the optimized circuit (similar to Figure \ref{fig:qnn}e) we observe that although the circuit classifies the original XOR data set poorly, it perfectly classifies a shifted version of the data described (Figure \ref{fig:qnn}a). This shows that we can learn new information by actually running the circuit (either on hardware or on simulator) and these new information can help us improve the use of quantum circuits.

\section{Discussion and Conclusion\label{sec:conclusion}}
% \hs{Edit based on new motivations. Pedagogical, Discuss limitations of the framework?}
% \hs{algo2qpu ... tool for testing circuits and discard or keep, }
With steady improvement of quantum hardware, coupled with developments in various software packages and cloud access to quantum processors, we are becoming better-equipped to test small instances of algorithms and perform important benchmark studies, ultimately to anticipate and prepare tasks for large-scale error-corrected quantum computers. In this paper, we have introduced a modular framework to guide and streamline 
% \pj{(how about ``guide and streamline'')}
the prototyping process for AHQC algorithms and demonstrated its use in designing and executing experiments by leveraging cloud access to quantum processors. Changes and improvements in each component of $\texttt{algo2qpu}$ will, in principle, benefit the overall framework and yield
a better pipeline for demonstrating the utility of quantum devices.\footnote{We note that Rigetti Computing has recently released an update to its platform, called the Quantum Cloud Services. Despite the changes, major steps of \texttt{algo2qpu} can still be applied and implemented using the updated code and service.}
% However, we note that the framework is limited by its assumption of the circuit model, that an algorithmic realization corresponds to the manipulation of collection(s) of quantum circuits.
% In the future, we may develop algorithms using alternative or higher level routines, in contrast to our current use of gate operations.
% \pj{(I vaguely know what you mean here... maybe worth clarifying though. What is meant by ``the framework is limited by its assumption of the circuit model''? Maybe there is a more positive way to present this idea? For example, ``While the current framework refers to adapting circuits, one could imagine extending this framework to adaptively adjust parameters of higher-level processes which implicitly involve lower-level quantum circuits.'' or something like that?)}
While the current framework refers to adapting circuits, e.g. directly tuning the gate parameters, one could imagine improving and extending this framework to adaptively adjust parameters of higher-level processes or routines which implicitly involve lower-level quantum circuits.
% \jr{Wait, Peter, Hannah, I don't think I understand the previous sentence. What are you trying to say here?}
% In addition, $\texttt{algo2qpu}$ relies on cloud-based quantum computing \pj{(Is it true that it relies on cloud-based quantum computing? E.g. I imagine that Rigetti could employ algo2qpu in-house. If not, maybe delete this sentence?)} and thus, the performance of the algorithm depends on the performance of the hardware provided by the service.
Nevertheless, continuing the efforts to build similar frameworks will enable efficient algorithmic testing that can eventually scale up to larger experiments and also provide a testbed for developing new and exploratory classes of algorithms, such as HQC algorithms in quantum machine learning. Just as AHQC algorithms were intended to make the most out of existing hardware, developing and/or formalizing software infrastructures that can unify multiple platforms, such as \texttt{algo2qpu}, will allow us to leverage capabilities of existing or near-term quantum software and hardware, setting the stage for more practical and powerful quantum computations. 

\begin{acknowledgments}
We thank Rigetti Computing for providing access to their quantum computer. We especially thank Ryan Karle for his help in setting up and running jobs on the 8Q-Agave device. The views expressed in this paper are those of the authors and do not reflect those of Rigetti Computing.
We also thank Jonathan Olson, Morten Kjaergaard, Max Radin, and Timothy Hirzel for helpful discussions and comments on the manuscript.
S. S. is supported by the DOE Computational Science Graduate Fellowship under grant number DE-FG02-97ER25308.
\end{acknowledgments}

\end{document}